\documentclass[12pt]{article}
\usepackage{amssymb, amsmath, amsthm, mathtools, amsbsy}

\usepackage{multirow}

\renewcommand{\hat}{\widehat}

\usepackage{natbib}
\usepackage{fullpage}

\usepackage[colorlinks,citecolor=blue,urlcolor=blue,linkcolor=red,naturalnames=true]{hyperref}
\usepackage{amsmath}
\usepackage[table]{xcolor}
\newcolumntype{L}{>$l<$}

\usepackage{todonotes}
\theoremstyle{plain}

\newtheorem{theorem}{Theorem}

\usepackage{algorithm,algorithmic}
\usepackage{setspace}
\usepackage{blindtext}
\usepackage{tabularx}
\usepackage{booktabs}
\usepackage{tocloft}
\usepackage{xr}
\usepackage{authblk}  
\usepackage{geometry}

\begin{document}

\title{\textbf{A Bayesian Nonparametric Approach for Semi-Competing Risks with Application to Cardiovascular Health}}

\author[1]{Karina Gelis-Cadena}
\author[1]{Michael Daniels}
\author[2]{Juned Siddique}

\affil[1]{University of Florida}
\affil[2]{Northwestern University}

\maketitle
\begin{abstract}
We address causal estimation in semi‐competing risks settings, where a non‐terminal event may be precluded by one or more terminal events. We define a principal‐stratification causal estimand for treatment effects on the non‐terminal event, conditional on surviving past a specified landmark time. To estimate joint event‐time distributions, we employ both vine‐copula constructions and Bayesian nonparametric Enriched Dirichlet‐process mixtures (EDPM), enabling inference under minimal parametric assumptions. We index our causal assumptions with sensitivity parameters. Posterior summaries via MCMC yield interpretable estimates with credible intervals. We illustrate the proposed method using data from a cardiovascular health study.
\end{abstract}
\section{Introduction}
Semi-competing risks data arise in settings where a non-terminal event may be censored by a terminal event, but not vice versa. These data structures frequently occur in longitudinal clinical studies, especially in the context of chronic diseases in which both disease progression and death are of interest. Unlike traditional competing risks, where all events are terminal and mutually exclusive, semi-competing risks account for a hierarchical dependency between events: the non-terminal event (e.g., hospitalization for heart failure (HF)) can only occur before the terminal event (e.g., death), while the terminal event can happen with or without the prior occurrence of the non-terminal event \citep{fine2001semi}.

In cardiovascular health research, this framework is particularly valuable. Patients with cardiovascular disease (CVD) often experience intermediate outcomes such as myocardial infarction, stroke, or hospitalization, which can be followed by death. Understanding the effect of treatment on these events, while properly accounting for the informative nature of death, is critical for making sound clinical decisions. Ignoring the semi-competing nature of such outcomes can lead to biased inferences, especially when the occurrence of death is related to the same underlying risk processes as the non-terminal event.

Recent approaches to semi-competing risks often employ the illness-death multistage model and copula-based methods for parameter estimation. Additionally, advancements have been made in formal causal analysis of semi-competing risk data to assess the causal effect of treatment on non-terminal events.\\

The illness-death model framework focuses on characterizing transition probabilities between health states, such as from healthy to diseased or from diseased to deceased, over time \citep{andersen2012statistical}. Frailty-based distributions are commonly incorporated within these models to account for unobserved heterogeneity and induce correlations between events. Recent advancements include applications to nested case-control designs, spline-based estimation techniques, penalized high-dimensional modeling, and Markov marginal structural approaches \citep{jazic2020estimation, huang2022nonparametric, reeder2023penalized, zhang2024marginal}.

Copula-based methods focus on the joint survival function's identifiable region, particularly when the terminal event occurs after the non-terminal event \citep{fine2001semi}. Recent advancements in this area have introduced models that incorporate flexible baseline hazard functions, various dependence structures, and regression frameworks to account for covariates. These methods have handled interval censoring, left truncation, and tail dependence, by including non-convex penalization and used frequentist approaches for causal interpretation and sensitivity analysis  \citep{wu2020meta, sorrell2022estimating, wei2023bivariate, sun2023semiparametric, sun2024penalised, yu2024exploring}.\\

In the context of formal causal analysis, two principal approaches are frequently employed to address semi-competing risks: mediation analysis and principal stratification. Mediation analysis \citep{baron1986moderator} decomposes the effect of an intervention on the primary outcome into two components: the indirect effect, which is mediated through a mediator, and the direct effect, which represents the effect not through the mediator. \citet{huang2021causal} formulated the semi-competing risks problem as a causal mediation analysis with the mediator and the primary outcome being non-terminal and terminal events, respectively. The direct effect represented the treatment effect directly on the terminal event while the indirect effect represented the treatment impact mediated by the non-terminal event. On the other hand, within a multi-state modeling \citet{valeri2021multistate} established non-parametric conditions to quantify the impact of stochastic interventions on non-terminal events that occur along the pathway between an exposure and a terminal event. Moreover, \citet{deng2024direct}  decomposed the total effect into a direct effect and an indirect effect under in completely randomized experiments by adjusting the prevalence and hazard of non-terminal events.\\ 

In principal stratification, introduced by \citet{frangakis2002principal}, the estimand is defined for a subpopulation classified by the joint outcomes of non-terminal events under both treatment and control conditions.
The Survivor Average Causal Effect (SACE) estimand is defined to compare potential outcomes among individuals who would survive under both treatment conditions. Building on this idea, \citet{xu2022bayesian} introduced a time-varying version of the SACE to assess the causal effect of treatment on a non-terminal event in the context of a randomized trial and developed a Bayesian nonparametric method for modeling the distribution of observable data. More recently, \citet{comment2025survivor} extended the SACE as a time-varying estimand to quantify the causal effect within the stratum of individuals who would have survived under both treatment conditions at a specified time point. Building on this framework, principal strata can be defined to focus on individuals susceptible to an intermediate event regardless of treatment \citep{gao2020defining} and further refined by stratifying subjects based on illness and death sequences, incorporating bivariate frailty models to account for heterogeneity \citep{nevo2022causal}. \\

In this study we will propose a Bayesian nonparametric (BNP) approach to evaluate the causal effect of treatment in a cohort study where a non-terminal event may be censored by up to two terminal events.

\section{D-Vine Distribution}

For $X_1,\dots, X_d$ a set of variables with joint distribution $F$ and density $f$, the joint distribution can be decomposed as 
\begin{align}\label{eq:joinDist}
f(x_1,\dots,x_d) &= f(x_d\mid x_1,\dots,x_{d-1})f(x_1,\dots,x_{d-1}) \notag \\
&= \cdots = \prod _{l=2}^d 
f(x_l\mid x_1,\dots,x_{l-1})\cdot f(x_1).
\end{align}

$f(x_l\mid x_1,\dots ,x_{l-1})$ can decomposed recursively as
\begin{align}
    f(x_l\mid x_1,\dots ,x_{l-1}) &= c_{1,l\mid 2,\dots ,l-1} \cdot f(x_l \mid x_2,\dots,x_{l-1}) \notag \\
    &= \left(\prod _{s=1}^{d-2} c_{
s,l\mid s+1,\dots,l-1}\right)·c_{(l-1),l} \cdot f_l(x_l),
\end{align}

\noindent where, $c(\cdot)$ denotes a bivariate copula density function. Specifically, \( c_{s,l \mid s+l,\dots,l-1} \) represents the conditional copula density between \( X_s \) and \( X_l \) given \( X_{s+l}, \dots, X_{l-1} \).\\

Thus, the joint density $f(x_1,\dots, x_d)$ can be written as:  

\begin{equation}\label{eq:dvine dist}
f(x_1, \dots, x_d) = \Bigl[\prod_{j=1}^d f_j(x_j)\Bigr]
\;\times\;\prod _{l=1}^ {d-1} \prod _{i=1} ^ {p-l} c_{i, (i+l) \mid (i+1), \dots,(i+l-1)}.  
\end{equation}

This representation decomposes the joint density $f(x_1,\dots , x_d)$  on marginals and pair copula densities, which are evaluated at conditional distribution functions. \citet{bedford2001probability, bedford2002vines} introduced decomposition (\ref{eq:dvine dist}) as a \textit{D-vine distribution}.\\

The following D-vine tree ilustrates the case where $d=4$

\begin{figure}[H]
  \centering
\begin{center}
\begin{tikzpicture}[
    node distance=1.5cm and 2.5cm,
    every node/.style={draw, rectangle, minimum size=0.8cm},
    every label/.style={font=\small, draw=none, fill=none},
    edge/.style={draw, -stealth},
]

\node (n1) {1};
\node[right=of n1] (n2) {2};
\node[right=of n2] (n3) {3};
\node[right=of n3] (n4) {4};

\path[edge] (n1) -- node[midway, below, draw=none] {12} (n2);
\path[edge] (n2) -- node[midway, below, draw=none] {23} (n3);
\path[edge] (n3) -- node[midway, below, draw=none] {34} (n4);

\node[draw=none, right=0.5cm of n4] {$T_1$};
\end{tikzpicture}
\end{center}
\begin{center}
\begin{tikzpicture}[
    node distance=1.5cm and 2.5cm,
    every node/.style={draw, rectangle, minimum size=0.8cm},
    every label/.style={font=\small, draw=none, fill=none},
    edge/.style={draw, -stealth},
]

\node (n12) {12};
\node[right=of n12] (n23) {23};
\node[right=of n23] (n34) {34};

\path[edge] (n12) -- node[midway, below, draw=none] {13$|$2} (n23);
\path[edge] (n23) -- node[midway, below, draw=none] {24$|$3} (n34);

\node[draw=none, right=0.5cm of n34] {$T_2$};
\end{tikzpicture}
\end{center}

   \centering
\begin{tikzpicture}[
    node distance=1.5cm and 2.5cm,
    every node/.style={draw, rectangle, minimum size=0.8cm},
    every label/.style={font=\small, draw=none, fill=none},
    edge/.style={draw, -stealth},
]
\node (n132) {13$|$2};
\node[right=of n132] (n243) {24$|$3};
\path[edge] (n132) -- node[midway, below, draw=none] {14$|$23} (n243);

\node[draw=none, right=0.5cm of n243] {$T_3$};
\end{tikzpicture}
\caption{D-vine representation of the joint density function  $f(x_1, x_2, x_3, x_4)$.}\label{D-vine_gen}
\end{figure}
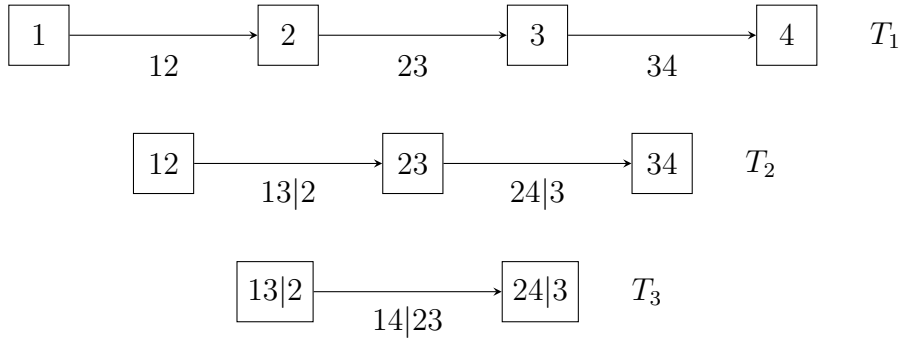

A D-vine decomposition provides a systematic framework for constructing a joint distribution in a stepwise manner: we begin with each identifiable marginal model and then introduce only the necessary bivariate conditional terms to capture dependence. In a causal setting with unobservable counterfactual pairs, this approach allows us to (i) estimate each marginal under standard identifiability arguments, (ii) include a single bivariate conditional component to model the unobserved dependence, and (iii) associate each identifying assumption with a specific edge in the vine.

\section{Proposed Semi-Competing Risks Causal Estimand with One Terminal Event and Identification Assumptions}\label{sect.One.Terminal.Events}

Let $Y_P^z$, $Y_D^z$ and $C^z$ denote progression time, death time, and censoring time, under treatment $z$. Here, $z = 0, 1$ represents control and treatment group, respectively. 
Fundamental to the setting is that $Y_P^z \ngtr Y_D^z$ (i.e., progression cannot happen after death).\\

\citet{xu2022bayesian} developed a Bayesian nonparametric (BNP) approach to assess the causal effect of treatment in a randomized trial where a non-terminal event may be censored by a terminal event, but not the reverse. Using the framework of principal stratification, they defined the estimand $\tau(\cdot)$ to capture the causal effect of interest as the function

\begin{equation}\label{estimand}
    \tau(u)=\frac{\Pr(Y_P^1 < u \mid Y_D^0 \geq u, Y_D^1 \geq u )}{\Pr(Y_P^0 < u \mid Y_D^0 \geq u, Y_D^1\geq u)}
\end{equation}
\noindent  where $\tau(\cdot)$ is a smooth function of $u$.\\

We consider the following D-vine representation of the joint density function  $f(Y_P^1, Y_D^1, Y_D^0, Y_P^0)$  \citep{ bedford2002vines}

\begin{figure}[H]
  \centering
    \centering
    \begin{tikzpicture}[grow=right, level distance=2.5cm, 
      every node/.style={draw, fill=blue!10, rectangle, inner sep=3pt, minimum size=0.8cm}]
      \node {$Y_P^1$}
         child {node {$Y_{D}^1$} 
           child {node {$Y_{D}^0$} 
             child {node {$Y_{P}^0$}}}};
    \end{tikzpicture}
  \vspace{1em} 

    \centering
    \begin{tikzpicture}[grow=right, level distance=3cm, 
      every node/.style={draw, fill=blue!10, rectangle, inner sep=1pt, minimum size=0.8cm}]
      \node {$Y_P^1,\,Y_{D}^1$} 
         child {node[fill=green!25]{$Y_{D}^1,\, Y_{D}^0$} 
           child {node {$Y_{P}^0,\,Y_{D}^0$}}};
    \end{tikzpicture}

  \vspace{1em} 


    \centering
    \begin{tikzpicture}[grow=right, level distance=4cm, 
      every node/.style={draw, fill=blue!10, rectangle, inner sep=1pt, minimum size=0.8cm}]
      \node[draw, fill=green!25] {$Y_P^1,\, Y_{D}^0 \mid Y_{D}^1$} 
         child {node[fill=green!25]{$Y_{P}^0,\, Y_{D}^1\mid Y_{D}^0$}};
    \end{tikzpicture}

  \vspace{1em} 

    \centering
    \begin{tikzpicture}[grow=right, level distance=4cm, 
      every node/.style={draw, fill=orange!25, rectangle, inner sep=1pt, minimum size=0.8cm}]
      \node {$Y_P^1,\, Y_{P}^0 \mid Y_{D}^1,\, Y_{D}^0$};  
    \end{tikzpicture}
  \caption{D-vine representation of the joint density function  $f(Y_P^1, Y_D^1, Y_D^0, Y_P^0)$.}\label{D-vine1}
\end{figure}
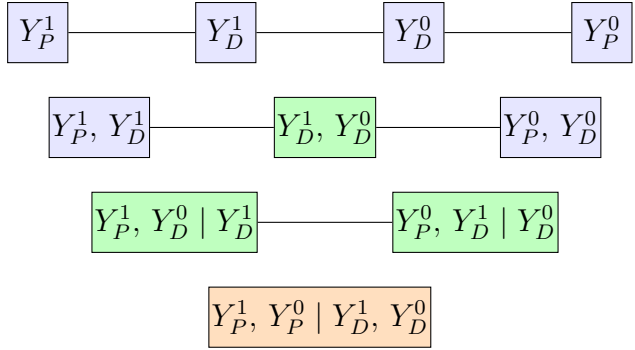

In the depicted vine tree diagram, the blue segments represent the distributions that can be identified from the observed data, the green segments correspond to the unidentified distributions, and the orange segment represents the distribution will not be necessary to identify (\ref{estimand}).
    
\section{Observed Data}

Let $Z$ denote treatment assignment and $\mathbf{X}$ denote a vector of the baseline covariates. Let $Y_P = Y_P^Z$, $Y_D = Y_D^Z$,
and $C = C^Z$. Let $T_1 = Y_P \wedge Y_D \wedge C$, $\delta = I(Y_P < Y_D \wedge C)$, $T_2 = Y_D \wedge C$, and $\xi = I(Y_D < C)$ denote the
observed event times and event indicators. The observed data for each observation are $O = (T_1, T_2, \delta, \xi, Z, \mathbf{X})$.
We assume that we observe $n$ i.i.d. copies of $O$.\\ 

\subsection{Identification Assumptions}\label{Assumptions One death}

We introduce the following four assumptions that are sufficient for identifying our causal estimand.\\

\noindent \textbf{Assumption 1:}  No unmeasured
confounders (NUC). Given a set of observed covariates $\mathbf{X=x}$, there are no unmeasured variables that confound the relationships between the treatment $Z$ and the variables $Y_p^z$,  $Y_{D}^z$, and $C^z$. This can be expressed as: 

$$(Y_P^z, Y_{D}^z, C^z) \perp Z \mid \mathbf{X=x}.$$

\noindent \textbf{Assumption 2:}  Censoring is non informative in the sense that
$$C^z \perp (Y_P^z, Y_D^z)  \mid (Z=z, \mathbf{X}= \mathbf{x}).$$

Let $\lambda_\mathbf{x}^z(t)$ and $ G_\mathbf{x}^z(t)$ denote the conditional hazard function and conditional distribution function of $Y_D^z$ given
$\mathbf{X} = \mathbf{x}$, respectively. Under Assumptions 1 and 2, $\lambda_\mathbf{x}^z(t)$ and $ G_\mathbf{x}^z(t)$ are identified via the following formula:

\begin{equation}
    \lambda_\mathbf{x}^z(t)= \lim _{dt \rightarrow 0} \left\{ \frac{\Pr(t \leq T_2 < t+ dt, \xi=1 \mid T_2 \geq t, \mathbf{X}=\mathbf{x}, Z=z]}{dt}\right \},
\end{equation}

\begin{equation}
    G_\mathbf{x}^z(t)=1- exp\left\{ -\int _0^t \lambda_x^z(s) ds \right\}.
\end{equation}

Furthermore, the conditional subdistribution function of $Y_P^z$ given $Y_D^z$ and $\mathbf{X} = \mathbf{x}$, $V_\mathbf{x}^z$, is identified via the following formula:

\begin{equation}
    V_\mathbf{x}^z(s \mid t) = \Pr(T_1 \leq s, \delta=1 \mid T_2=t, \xi =1, \mathbf{X}=\mathbf{x}, Z=z].
\end{equation}

\noindent \textbf{Assumption 3:} The conditional joint distribution function of $(Y_D^1, Y_D^0)$ given $X = x$, $G_\mathbf{x}$, follows a Gaussian
copula model, i.e.,

\begin{equation}\label{copula 1}
 G_\mathbf{x}(v,w ;\rho)= \Phi_{2,\rho}\Big[\Phi^{-1}\left\{G_\mathbf{x}^1(v)\right\}, \Phi^{-1}\left\{G_\mathbf{x}^0(w)\right\}\Big],   
\end{equation}

\noindent where $\Phi$ is is the univariate standard normal CDF and $\Phi_{2,\rho}$ is  the bivariate normal CDF with mean 0, marginal variances
1, and correlation $\rho \in (-1, 1)$. For fixed $\rho$, $G_\mathbf{x}$ is identified since $G_\mathbf{x}^1$ and $G_\mathbf{x}^0$ are identified.\\

To identify the causal estimand, for one assumption \citet{xu2022bayesian} assumed that the progression time under treatment $z$ is conditionally independent of the death time under treatment $1-z$ given the death time under treatment $z$ and covariates $\mathbf{X=x}$. Mathematically, this is expressed as $Y_P^z \perp Y_D^{1-z} \mid Y_D^z, X=x$. While this assumption simplifies the analysis, it could be too restrictive in practice; we aim to provide greater flexibility by introducing a more flexible assumption.\\

\noindent \textbf{Assumption 4:}  (New) The conditional joint distribution function of $(Y_P^z, Y_D^{1-z})$ given $(Y_D^z =t, X = x)$, $H^z_x$ for $z=0,1$ follows a Gaussian
copula model, i.e.,
\begin{equation}\label{copula 2}
 H_\mathbf{x}^z(s,r \mid t ;\rho^*_z)= \Phi_{2,\rho^*_z}[\Phi^{-1}\left\{\Pr(Y_p^z \leq s \mid Y_D^z=t]\right\}, \Phi^{-1}\left\{\Pr(Y_D^{1-z} \leq r \mid Y_D^z = t]\right\}],  
\end{equation}

\noindent where $z=0,1$, $\Pr(Y_p^z \leq s \mid Y_D^z=t) =V_\mathbf{x}^z(s \mid t)$ and  $\Pr(Y_D^{1-z} \leq r \mid Y_D^z = t)$ is identified through $G_\mathbf{x}(r, t)$.

\subsection{Estimand Identification}

The probabilities in the estimand $\tau(u)$ are conditional on survival past the threshold $u$ in both death outcomes; therefore, we integrate only over the regions where $Y_D^1 \geq u$ and $Y_D^0 \geq u$. Denote by $dG_x(v,w)$ the joint density of the death times $(Y_D^1, Y_D^0)$ for a subject with baseline covariates $x$, and by $dK(x)$  the measure on the covariate space. Then, for a fixed treatment level  $z$, the conditional probability that the progression occurs before $u$ (given survival) is obtained by ``averaging" the conditional probabilities derived from the copula model. 
We build our estimator by integrating the joint measure: $\int _{s < u} dH_x^z(s, w \mid v)$, which represents the (infinitesimal) joint probability mass (given $Y_D^z$ and $x$) that $Y_P^z < u$ and that $Y_D^{1-z}$ is (approximately) equal to $w$.\\

Then, the overall conditional probability for treatment $z$  is obtained by integrating over the complementary death time $w$ (with $w \geq u$), over the anchor death time $v$ (with $v \geq u$), and finally over $x$. 

\begin{theorem} 
Under Assumptions 1-4 , $\tau(\cdot)$,  is identified from the distribution of the observed data as follows:
$$ \tau(u)= \frac{\int _x \int_{v \geq u} \int _{w \geq u}  \left[\int _{s< u}dH_\mathbf{x}^1(s,w \mid v) \right] dG_\mathbf{x}(v,w) dK(x)}{\int _x  \int_{v \geq u} \int _{w \geq u} \left [\int _{s< u} dH_\mathbf{x}^0(s,v \mid w)\right] dG_\mathbf{x}(v, w) dK(x)}. $$
\end{theorem}

\section{Proposed Semi-Competing Risks Model with Two Terminal Events}\label{sect.Two.Terminal.Events}

This extension to include two terminal events is motivated by the objective of estimating the causal effect of a treatment in a cohort study, where the occurrence of  non-terminal HF may be censored by death due to CVD or death from other causes, but not the reverse.\\

In this context, let $Y_P^z$, $Y_{D_1} ^z$, $Y_{D_2} ^z$ and $C^z$ represent the progression time (age at first non-terminal HF event), time at death due to CVD; and  time at death due to non-CVD, and censoring time under treatment $z$, respectively. Here, $z$ represents the medication status at baseline and, $z = 0, 1$ represents not on, or on, medication, respectively. \\

Fundamental to our setting is that $Y_P^z \ngtr Y_{D_1}^z$ and $Y_P^z \ngtr Y_{D_2}^z$ (i.e., progression cannot happen after death).\\ 

The principal strata we consider includes subjects who
\begin{enumerate}
    \item Survive beyond time $u$. 
    \item Under either medication status, experience death due to CVD ``before" death due to other causes.
\end{enumerate}

The principal strata, therefore, are defined by the pair

\begin{enumerate}
    \item $Y_{D_2}^1 \geq Y_{D_1}^1 \geq u$: subjects on medication at baseline, whose death occurs after time $u$ and it is due to CVD.
     \item $Y_{D_2}^0 \geq Y_{D_1}^0 \geq u$: subjects not on medication at baseline, whose death occurs after time $u$ and it is due to CVD.
\end{enumerate}

Thus, we define the following causal estimand of interest:

\begin{equation}\label{new-estimand}
    \tau(u)=\frac{\Pr(Y_P^1 < u \mid Y_{D_2}^1 \geq Y_{D_1}^1 \geq u, Y_{D_2}^0\geq  Y_{D_1}^0 \geq u )}{\Pr(Y_P^0 < u \mid Y_{D_2}^1 \geq Y_{D_1}^1 \geq u, Y_{D_2}^0\geq  Y_{D_1}^0 \geq u)}.
\end{equation}

The estimand $\tau(\cdot)$ compares the likelihood of the first non-terminal HF event occurring prior to time $u$ for a subject on medication at baseline relative to a subject not on medication at baseline, among patients who survive up to time $u$ and whose primary cause of death was CVD. \\

We need several assumptions to identify (\ref{new-estimand}). To facilitate this, we decompose the joint distribution $f(Y_P^1, Y_{D_1}^1, Y_{D_2}^1,  Y_{D_2}^0, Y_{D_1}^0, Y_P^0)$ using a D-vine tree structure as outlined below 

\begin{figure}[H]
  \centering
\begin{tikzpicture}[grow=right, level distance=2.5cm, every node/.style={draw,fill=blue!10, rectangle,inner sep=3pt,minimum size=0.8cm}]
  \node {$Y_P^1$}
   child {node{$Y_{D_1}^1$} child {node {$Y_{D_2}^1$} child {node {$Y_{D_2}^0$} child{node{$Y_{D_1}^0$} child{node{$Y_P^0$}}}}}};
       \end{tikzpicture}

\begin{center}
\begin{tikzpicture}[grow=right, level distance=3cm, every node/.style={draw,fill=blue!10, rectangle,inner sep=1pt,minimum size=0.8cm}]
     \node {$Y_P^1,Y_{D_1}^1$} child {node{$Y_{D_1}^1, Y_{D_2}^1$} child {node[fill=green!25] {$Y_{D_2}^1,Y_{D_2}^0$} child {node {$Y_{D_1}^0, Y_{D_2}^0 $} child{node{$ Y_P^0,Y_{D_1}^0$}}}}};
   \end{tikzpicture}
   \end{center}
     
\begin{center}
\begin{tikzpicture}[grow=right, level distance=4cm, every node/.style={draw,fill=blue!10,rectangle,inner sep=1pt,minimum size=0.8cm}]
     \node{$Y_P^1,  Y_{D_2}^1 \mid Y_{D_1}^1$} 
   child {node[fill=green!25]{$Y_{D_1}^1,  Y_{D_2}^0\mid Y_{D_2}^1$} child {node[fill=green!25] {$Y_{D_1}^0,Y_{D_2}^1 \mid Y_{D_2}^0$} child {node{$Y_P^0,Y_{D_2}^0\mid Y_{D_1}^0$}}}};
\end{tikzpicture}
\end{center}      
\begin{center}
   \begin{tikzpicture}[grow=right, level distance=4cm, every node/.style={draw,fill=green!25, rectangle,inner sep=1pt,minimum size=0.8cm}]
    \node {$Y_P^1,  Y_{D_2}^0 \mid Y_{D_1}^1, Y_{D_2}^1$} 
   child {node{$Y_{D_1}^1,  Y_{D_1}^0\mid Y_{D_2}^1, Y_{D_2}^0$} child {node {$Y_P^0,Y_{D_2}^1\mid Y_{D_1}^0, Y_{D_2}^0$}}};
\end{tikzpicture} 
\end{center}      
\begin{center}
    \begin{tikzpicture}[grow=right, level distance=6cm, every node/.style={draw,fill=green!25, rectangle,inner sep=1pt,minimum size=0.8cm}]
    \node {$Y_P^1,  Y_{D_1}^0 \mid Y_{D_1}^1, Y_{D_2}^1, Y_{D_2}^0$} 
   child {node{$ Y_P^0,Y_{D_1}^1\mid Y_{D_1}^0, Y_{D_2}^1, Y_{D_2}^0$}};
\end{tikzpicture}
    \end{center}
\begin{center}
 \begin{tikzpicture}[grow=right, level distance=7cm, every node/.style={draw,fill=blue!10, rectangle,inner sep=1pt,minimum size=0.8cm}]
   \node[fill=orange!20] {$Y_P^1,  Y_P^0 \mid Y_{D_1}^1, Y_{D_2}^1, Y_{D_1}^0, Y_{D_2}^0$} ;
\end{tikzpicture}   
\end{center}
 \caption{D-vine representation of the joint density function  $f(Y_P^1, Y_{D_1}^1, Y_{D_2}^1,  Y_{D_2}^0, Y_{D_1}^0, Y_P^0)$.}\label{D-vine2}
\end{figure}
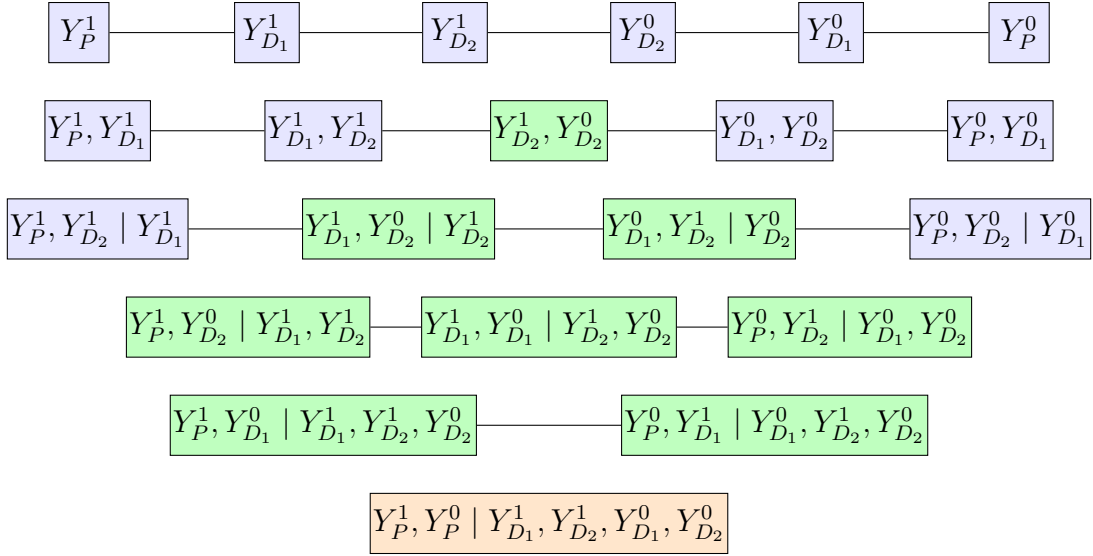

In the above, the blue segments represent the distributions identified from the observed data, the green segments correspond to the unidentified distributions, and the orange segment denotes the distribution that is not needed to identify (\ref{new-estimand}).
\subsection{Observed Data and Notation}

Let $Z$ denote treatment assignment and $\mathbf{X}$ denote a vector of the baseline covariates. Let $Y_P = Y_P^Z$, $Y_{D_1} = Y_{D_1}^Z$, $Y_{D_2} = Y_{D_2}^Z$, and $C = C^Z$. Let $T_1 = Y_P \wedge Y_{D_1} \wedge Y_{D_2} \wedge C$, $\delta = I(Y_P < Y_{D_1} \wedge Y_{D_2} \wedge C)$, $T_2 = Y_{D_1} \wedge Y_{D_2} \wedge C$, $\xi_1 = I(Y_{D_1} < Y_{D_2} \wedge C)$ and $\xi_2 = I(Y_{D_2} < Y_{D_1} \wedge C)$ denote the observed event times and event indicators. The observed data for each observation are $O = (T_1, T_2, \delta, \xi_1, \xi_2, Z, \mathbf{X})$.
We assume that we observe $n$ i.i.d. copies of $O$.\\

We use the following notation to define key variables, and functions used in the model and analysis.\\

\begin{itemize}
\item $\bullet$ $G_\mathbf{x}^ z $ (identified from the observed data) denotes the conditional distribution of $Y_{D_1}^z $ given $Y_{D_2}^z$  and $\mathbf{X=x}$ 
$$G_\mathbf{x}^ z (s \mid t) = Pr ( Y_{D_1}^z \leq s \mid Y_{D_2}^z = t, \mathbf{X=x}, Z=z), \quad \text{for} \quad t \geq s.$$

\item $\bullet$ $V_{\mathbf{x}}^z$ (identified from the observed data) denotes the conditional subdistribution function of $Y_p^z$ given $(Y_{D_1}^z, Y_{D_2}^z)$ and $\mathbf{X=x}$
\begin{align*}
  V_{\mathbf{x}}^z(r\mid s,t) &= \Pr(Y_P^z \leq r,  \mid Y_{D_1}^z= s,  Y_{D_2}^z=t)\\
  &= \Pr(T_1 \leq r, \delta=1 \mid T_2=\min(t,s), \mathbf{X=x}, Z=z),  
\end{align*}
where $r \leq \min( s, t)$.

\item $\bullet$ $\lambda_{\mathbf{x}}^z$ denotes the hazard function of the non-terminal event
\begin{equation}
    \lambda_{\mathbf{x}}^z(t)= \lim _{dt \rightarrow 0} \left\{ \frac{\Pr(t \leq T_1 < t+ dt, \delta=1 \mid T_1 \geq t, T_2 \geq t, \mathbf{X}=\mathbf{x}, Z=z)}{dt}\right \}.
\end{equation}

\item $\bullet$ $\lambda_{D_1}^z$ denotes the cause-specific hazard function of death due to CVD, i.e., the instantaneous risk of experiencing death due to CVD at time $t$, given that the individual has survived up to time $t$.
\begin{equation}
    \lambda_{D_1}^z(t)= \lim _{dt \rightarrow 0} \left\{ \frac{\Pr(t \leq T_2 < t+ dt, \xi_1=1 \mid T_2 \geq t, \mathbf{X}=\mathbf{x}, Z=z)}{dt}\right \}
\end{equation}

\item $\bullet$ $\lambda_{D_2}^z$ denotes the cause-specific hazard function of death due to non-CVD, i.e., the instantaneous risk of experiencing death due to non-CVD at time $t$, given that the individual has survived up to time $t$.
\begin{equation}
    \lambda_{D_2}^z(t)= \lim _{dt \rightarrow 0} \left\{ \frac{\Pr(t \leq T_2 < t+ dt, \xi_2=1 \mid T_2 \geq t, \mathbf{X}=\mathbf{x}, Z=z)}{dt}\right \}.
\end{equation}
\end{itemize}

\subsection{Identification Assumptions}\label{Sec. Assumptions two deaths}

We introduce the following six assumptions that are sufficient for identifying our causal estimand.

\noindent \textbf{Assumption 1:}  No unmeasured
confounders (NUC). Given a set of observed covariates $\mathbf{X=x}$, there are no unmeasured variables that confound the relationships between the treatment $Z$ and the variables $Y_p^z$,  $Y_{D_1}^z$, $Y_{D_2}^z$ and $C^z$. This can be expressed as: 
$$(Y_P^z, Y_{D_1}^z, Y_{D_2}^z, C^z) \perp Z \mid \mathbf{X=x}.$$

\vspace{6pt}
\noindent \textbf{Assumption 2:} Censoring is non informative in the sense that
$$C^z \perp (Y_P^z, Y_{D_1}^z, Y_{D_2}^z)  \mid (Z=z, \mathbf{X}= \mathbf{x}).$$

\vspace{6pt}
\noindent \textbf{Assumption 3:} The conditional joint distribution function of $(Y_{D_2}^1, Y_{D_2}^{0})$ given $\mathbf{X}=\mathbf{x}$,  $H_\mathbf{x}$ follows a Gaussian copula model, i.e.,
 \begin{equation*}
 H_\mathbf{x}(t,w)= \Phi_{2,\rho}\Big[\Phi^{-1}\left\{ \Pr(Y_{D_2}^1 \leq t)\right\}, \Phi^{-1}\left\{ \Pr(Y_{D_2}^{0} \leq w)\right\}\Big],  
\end{equation*} 

\noindent where $ \Pr(Y_{D_2}^1 \leq t)$ and $\Pr(Y_{D_2}^{0} \leq w)$ are identified from the observed data. \\
\noindent A single sensitivity parameter $\rho$ is required in this assumption.\\
 \vspace{6pt} 

\noindent \textbf{Assumption 4:} The conditional joint distribution function of $(Y_{D_1}^z, Y_{D_2}^{1-z})$ given $(Y_{D_2}^z =t, X=x)$, $J_\mathbf{x} ^{z}$, for $z = 0, 1$ follows a Gaussian copula model, i.e.,
\begin{equation*}
 J_\mathbf{x}^{z}(s,w \mid t) = \Phi_{2,\rho^*_z}\Big[\Phi^{-1}\left\{ \Pr(Y_{D_1}^z \leq s \mid Y_{D_2}^z=t)\right\}, \Phi^{-1}\left\{\Pr(Y_{D_2}^{1-z} \leq w  \mid Y_{D_2}^z=t)\right\}\Big],
\end{equation*} 

\noindent where  $\Pr(Y_{D_1}^z \leq s \mid Y_{D_2}^z=t)=G_\mathbf{x} ^z (s \mid t)$ and $\Pr(Y_{D_2}^{1-z} \leq w  \mid Y_{D_2}^z=t)$ can be identified through ${H_\mathbf{x}(t,w)}$ for $z=0,1$.\\
\noindent Two sensitivity parameters $\rho^*_z$, for $z=0,1$ are required in this assumption.\\
\vspace{6pt}

\noindent \textbf{Assumption 5:} Conditional cross independence for progression.\\
\noindent $Y_p^z$ is conditionally independent of $Y_{D_2}^{1-z}$ given $Y_{D_1}^z$, $Y_{D_2}^z$ and $\mathbf{X}= \mathbf{x}$, ie.,
\begin{equation*}
    Y_P^z \perp Y_{D_2}^{1-z} \mid Y_{D_1}^z, Y_{D_2}^z, \quad z=0,1.
\end{equation*} 

\noindent $Y_p^z$ is conditionally independent of $Y_{D_1}^{1-z}$ given $Y_{D_1}^z$, $Y_{D_2}^z$, $Y_{D_2}^{1-z}$ and $\mathbf{X}= \mathbf{x}$, ie.,
\begin{equation*}
   Y_P^z \perp  Y_{D_1}^{1-z} \mid Y_{D_1}^z, Y_{D_2}^z, Y_{D_2}^{1-z} \quad z=0,1.
\end{equation*} 
\vspace{6pt}
\noindent \textbf{Assumption 6:} The conditional joint distribution function of $(Y_{D_1}^1, Y_{D_1}^0)$ given $(Y_{D_2}^1=t, Y_{D_2}^0=w, \mathbf{X}=\mathbf{x})$, $L_\mathbf{x}$, follows a copula model, i.e.,
\small{
\begin{equation*}
 L_\mathbf{x}(s,v \mid t, w)= \Phi_{2,\rho^{**}}\Big[\Phi^{-1}\left\{ \Pr(Y_{D_1}^1 \leq s \mid Y_{D_2}^1=t, Y_{D_2}^0=w)\right\}, \Phi^{-1}\left\{ \Pr(Y_{D_1}^0 \leq v  \mid Y_{D_2}^1=t, Y_{D_2}^0=w)\right\}\Big],  
\end{equation*}}

\noindent where $\Pr(Y_{D_1}^z \leq s \mid Y_{D_2}^z=t, Y_{D_2}^{1-z}=w)$ can be identified through $J_\mathbf{x}^{z}(s, w \mid t)$ and $G_\mathbf{x}^z(s \mid w)$, $z=0,1$. \\
\noindent A single sensitivity parameter $\rho^{**}$ is required to specify the correlation in the copula distribution.\\

In this assumption we employ a Gaussian copula to unite the two event time distributions CVD and non-CVD death under both treatments arms. Because the two potential outcomes cannot be observed simultaneously, the copula reduces all assumptions about their counterfactual dependence to a single correlation parameter. This construction allows us to transparently evaluate how varying degrees of correlation affect our estimand, without imposing any additional structure on the marginal models.

\subsection{Estimand Identification}

Recall the estimand $\tau(u)$ defined in (\ref{new-estimand}),
$$\tau(u) = \frac{\Pr(Y_P^1 < u \mid Y_{D_2}^1 \geq Y_{D_1}^1 \geq u, Y_{D_2}^0\geq  Y_{D_1}^0 \geq u )}{\Pr(Y_P^0 < u \mid Y_{D_2}^1 \geq Y_{D_1}^1 \geq u, Y_{D_2}^0\geq  Y_{D_1}^0 \geq u)}.$$

The probabilities defining the estimand  are conditional on two related survival criteria. First, subjects must survive past the threshold $u$; and second, under either medication status, the subject must experience death due to CVD ``before" death from other causes. Thus, we integrate only over the regions where $Y_{D_1}^z\geq u$, $Y_{D_1}^{1-z}\geq u$, $Y_{D_2}^z\geq Y_{D_1}^z$, and $Y_{D_2}^{1-z}\geq Y_{D_1}^{1-z}$ for $z=0,1$. \\
 
 \noindent Thus the numerator of  $\tau(u)$ can be expressed as the integral of the joint density over those regions,
  \small{
\begin{equation*}
    \int_ \mathbf{x} \int_{r <u}\int _{s \geq u} \int _{t \geq s} \int _{v \geq u} \int _{w \geq v}   \Pr(Y_P^1=r \mid Y_{D_1}=s, Y_{D_2}^1=t, Y_{D_1}^0=v, Y_{D_2}^0=w)dL_{\mathbf{x}}(s,v \mid t, w) dH_\mathbf{x}(t,w)dK(\mathbf{x})
\end{equation*}}

\noindent where $dL_\mathbf{x}(s,v \mid t,w)$ and $dH_\mathbf{x}(t, w)$ are determined by Assumption 6 and Assumption 3 respectively.\\
 
Under Assumption 5 (conditional cross‐independence of progression), it follows that

$$\Pr(Y_P^1=r \mid Y_{D_1}=s, Y_{D_2}^1=t, Y_{D_1}^0=v, Y_{D_2}^0=w)=dV_\mathbf{x}^1(r \mid s,t).$$ 
Therefore, proceeding similarly for the denominator,  $\tau(\cdot)$ can be identified from the distribution of the observed data. 

\begin{theorem} 
Under Assumptions 1-6 , $\tau(\cdot)$,  is identified from the distribution of the observed data as follows:
\begin{equation*}
     \tau(u)=\frac{ \int_ \mathbf{x} \int _{r < u} \int _{s \geq u} \int _{t \geq s} \int _{v \geq u} \int _{w \geq v} dV_{\mathbf{x}}^1(r \mid s, t) dL_{\mathbf{x}}(s,v \mid t, w) dH_\mathbf{x}(t,w)dF(\mathbf{x}) }{ \int_ \mathbf{x} \int _{r < u}\int _{s \geq u} \int _{t \geq s} \int _{v \geq u} \int _{w \geq v} dV_{\mathbf{x}}^0(r \mid v,w) dL_{\mathbf{x}}(s,v \mid t, w) dH_\mathbf{x}(t,w) dF(\mathbf{x})}.
\end{equation*}
\end{theorem}

\section{Observed Data Models}

\subsection{Enriched Dirichlet Process Mixture}
When modeling the joint density of the response and covariates $(Y, X)$ as a DPM with a large number of predictors, difficulties arise. Subjects with similar covariates tend to cluster together, causing clusters to be primarily based on covariate similarity, which results in poor estimates \citep{wade2011enriched}. The enriched Dirichlet process mixture (EDPM) is a conjugate nonparametric prior that extends the DPM by enabling a nested clustering structure, where the top level $y$-clusters are based on the regression of the response on the predictors and within each $y$-cluster there
are bottom level $x$-clusters based on the predictors \citep{wade2014improving}. This structure addresses the challenges of jointly modeling $(Y,X)$, resulting in improved predictions.
\begin{align*}
    [Y_i \mid X_i, \theta_i ]& \sim p(y \mid x,\theta_i)\\
    [X_{i,j} \mid \omega _i] &\sim p(x_j \mid \omega _i)\\
    [(\theta_i, \omega_i) \mid F] & \sim F \\
    F & \sim EDP(\alpha_\theta, \alpha_\omega, H).
\end{align*}

\noindent The term $F \sim EDP(\alpha_\theta, \alpha_\omega, H)$ means that $F(d\theta, d\omega)= F_\theta(d\theta) \times F_{\omega \mid \theta}(d\omega \mid \theta)$ with $F_\theta \sim DP(\alpha _\theta,H_\theta)$, $F_{\omega \mid \theta} \sim DP(\alpha_\omega,H_{\omega \mid \theta})$ and $H=H_\omega \times H_{\omega \mid \theta}$. 

Analogous to the stick-breaking construction for the DP, the joint distribution of $(Y,X)$ can be represented using a squared breaking construction given by
\begin{equation}\label{eq:EDPM}
    p(y; \theta) = \sum_{k=1}^\infty  \gamma_k p(y\mid x; \theta) \sum_{j=1}^\infty \gamma_{j\mid k} p(x; \omega_{j\mid k}), 
\end{equation}
where 
\begin{equation*}
   \gamma_k =\nu_k\prod _{l < k} (1 - \nu _l), \quad \nu_l \sim Beta(1, \alpha_\theta), \quad \theta_k  \overset{\text{iid}}{\sim} H_\theta
\end{equation*}
\begin{equation*}
   \gamma_{j\mid k} =\nu_{j\mid k}\prod _{l < j} (1 - \nu _{l \mid k}), \quad \nu_{l\mid j} \sim Beta(1, \alpha_\omega), \quad \omega_{j\mid k}  \overset{\text{iid}}{\sim} H_{\omega \mid \theta}
\end{equation*}

 \citet{burns2023truncation} propose a truncated version of the EDPM, where (\ref{eq:EDPM}) is rewritten as:
 
\begin{equation}\label{eq:Trun EDPM}
    p(y; \theta) = \sum_{k=1}^N \sum_{j=1}^M  \gamma_k   \gamma_{j\mid k} p(y\mid x; \theta) p(x; \omega_{j\mid k}), 
\end{equation}

\noindent where $\gamma_1 =\nu_1$,  $\gamma_k =\nu_k\prod _{l=1}^{k-1} (1 - \nu _l)$, $k=2,\dots, N$; $\gamma_{1\mid k} =\nu_{1\mid k}$, $\gamma_{j\mid k} =\nu_{j\mid k}\prod _{l=1}^{j-1} (1 - \nu _{l \mid k})$, $k = 1, \dots, N$  and $j = 2, \dots,M$; $\nu_k \sim Beta(1, \alpha_\theta)$, $k=1, \dots, N-1$; $\nu_N=1$; and for  $k=1, \dots, N$  $\nu_{j\mid k} \sim Beta(1, \alpha_\omega)$ , $j=1, \dots, M-1$, and $\nu_{M\mid k}=1$. This truncation facilitates a simple blocked Gibbs sampler for posterior computation (see Appendix \ref{appendix A}).

\section{Cardiovascular Health Data Analysis}

We used the proposed causal estimand on Section \ref{sect.One.Terminal.Events} and \ref{sect.Two.Terminal.Events} to estimate the causal effect of treatment in a cohort study where the occurrence of the first non-terminal HF event, may be censored by two terminal events: death due to CVD and death due to other causes, but not vice versa. HF is considered non-terminal if the subject does not die within 30 days following the HF event. Additionally, we focus on a baseline age range between 40 and 59 years. The analysis aims to understand how the treatment impacts the timing of the non-terminal HF event while accounting for the possibility that these terminal events can prevent its observation.\\

The data used in this analysis come from the Framingham Heart Study (FHS), a longitudinal cohort study designed to investigate risk factors for CVD. We restrict our attention to participants who were free of CVD at baseline, specifically excluding individuals with a history of coronary heart disease. The final analytic sample consists of 1833 individuals.

Each individual is characterized by a set of nine baseline covariates measured at the start of the study: systolic blood pressure, total cholesterol, high-density lipoprotein cholesterol, age, sex, smoking status, diabetes status, race, and whether the participant was receiving treatment for hypertension. 

In addition to these covariates, we consider three time-to-event outcomes. First, the time to non-terminal HF, recorded as the number of years from baseline to the first diagnosis of HF. Second, the time to death due to CVD, and third, the time to death from any cause. For each of these events, we define binary indicators to denote whether the event occurred during the follow-up period. Specifically, we define an indicator for incident HF, an indicator for death due to CVD, and an indicator for death due to any cause. Censoring occurs when a participant does not experience the event of interest during the observed follow-up time.

All individuals were followed prospectively from their baseline visit until the earliest of the following: death, loss to follow-up, or the end of the study period. Table \ref{tab:FHS} summarizes the joint distribution of incident HF status and vital status (CVD‐related death, non‐CVD death, or alive) among the 1833 Framingham Heart Study participants.

\begin{table}[htbp]
  \caption{Cross‐classification of incident HF status and vital status (CVD‐related death, non‐CVD death, or alive) among 1833 FHS participants} \label{tab:FHS}
    \centering
      \begin{tabularx}{\textwidth}{XXXXX}
    \hline
    & {CVD Dead} & {Non-CVD Dead} & {Alive} &{Total}\\ \hline
{HF}& 207 & 160 & 33 & 400    \\
{Non-HF}& 278 & 973 & 182 &  1433  \\  
{Total} & 485 & 1133 & 215 & 1833\\ \hline
\end{tabularx}
\end{table}

\begin{figure}[H]
 \centering
\includegraphics[scale=0.55]{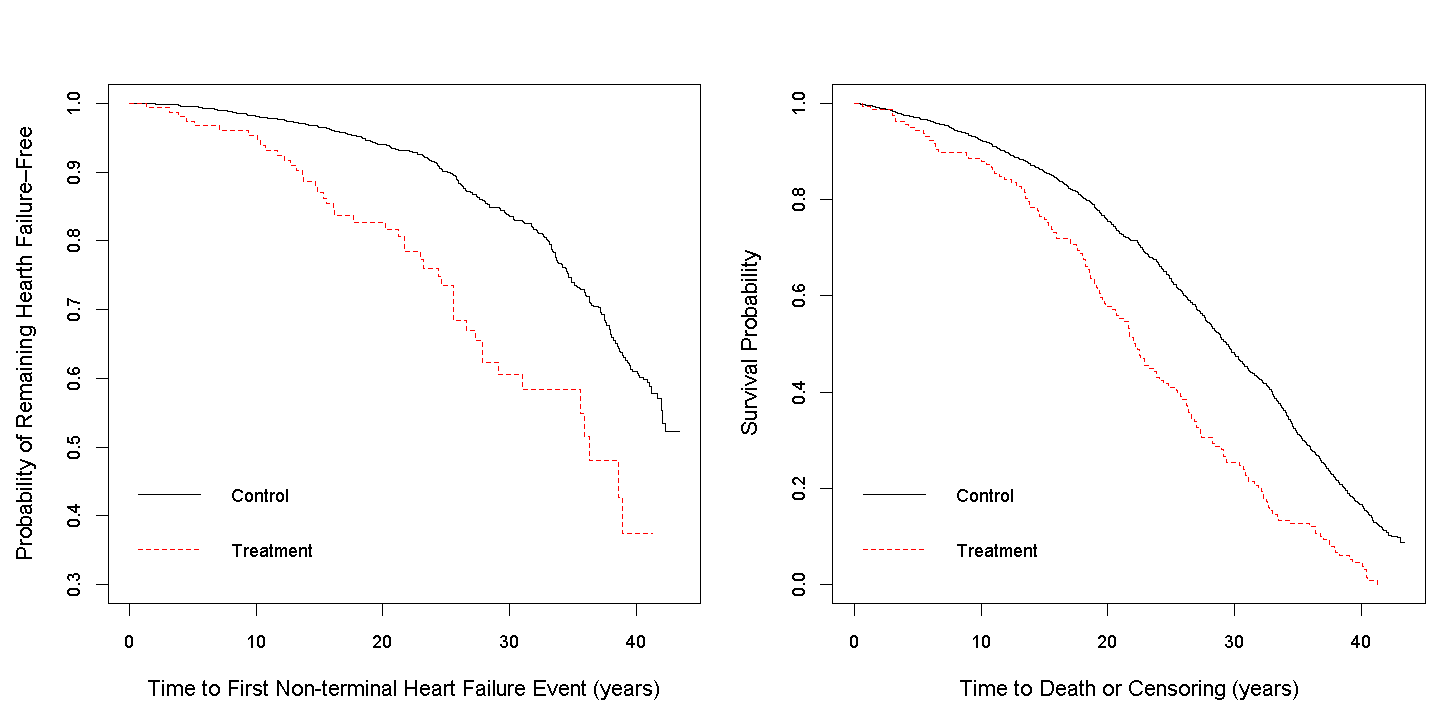}
\caption{Kaplan–Meier estimates of HF-free survival (left) and overall survival (right) stratified by hypertension treatment status.}\label{KM-HF-DTH}
\end{figure}

Figure \ref{KM-HF-DTH} displays side by side Kaplan-Meier curves for observed HF free time and the observed survival data by hypertension treatment status. In both panels the survival function for the treated group lies consistently below that of the control group, reflecting a higher cumulative incidence of HF and shorter overall survival among individuals receiving treatment. Because treatment was not randomized, this contrast likely reflects confounding by indication: patients who are prescribed medication tend to have more severe hypertension or other baseline risk factors, and thus experience events earlier than untreated individuals. We perform a causal analysis to examine this. 

\subsection{Results for the One Terminal Event}

Posterior inference was performed using the NIMBLE R package, which provides a flexible system for specifying hierarchical models and implementing MCMC algorithms. NIMBLE uses adaptive Metropolis, within-Gibbs sampling by default, allowing efficient sampling from complex posterior distributions \citep{de2017programming}. Posterior inferences were obtained using the EDPM truncation approximation (with $N=10$ and $M=8$), based on 40000 MCMC iterations with a burn in period of 20000 iterations; we set the concentration parameter $\alpha_\theta=1$ and assumed 
$\alpha_{\omega \mid \theta} \sim Gamma(1,1)$. In Appendix \ref{appendix model specification} we provide more information about the priors. \\

Table \ref{estimand1-onedeath} and \ref{estimand2-onedeath} presents the estimated values of the estimand $\tau(u)$ (defined in \ref{estimand}), for still being alive $u$ years after enrollement, $u=10, \, 20, \,30$, and 
40 years, under various specifications of the copula correlation parameters. For both baseline correlations $\rho=0.3$ and 0.6 (Assumption 3)  and  both conditional correlation scenarios $\rho^*_z=0.3$ and 0.6 (Assumption 4), the posterior means of $\tau(u)$ remain close to unity. The $95\% $ credible intervals for every scenario covers 1, indicating substantial posterior uncertainty. In particular, although the upper bounds exceed 1, suggesting the possibility that treated patients may experience a higher HF rate, the fact that the intervals also dip below 1 means we cannot confidently rule out either a protective or harmful effect. 

\begin{table}
\caption{Posterior estimates of $\tau(u)$ for correlation $\rho=0.3$ (Assumption 3)  under two scenarios using correlation values $\rho^*_z=0.3,0.6$ (Assumption 4) for $z=0,1$, with 95$\%$ point-wise credible intervals.} \label{estimand1-onedeath}
\begin{tabularx}{\textwidth}{XXX}
\toprule
Threshold & $\rho_z^*=0.3$ &  $\rho_z^*=0.6$ \\
\midrule
 $u=10$ & 0.86  \; (0.53, \, 1.26) &  0.80 \; (0.45, \, 1.27)\\
 $u=20$ & 0.93 \; (0.68, \, 1.18) &  0.93 \; (0.63, \, 1.28) \\
 $u=30$ & 0.96 \; (0.75, \, 1.18) & 1.01 \; (0.69, \, 1.41) \\
 $u=40$ & 0.94 \; (0.75, \, 1.13) & 0.95 \; (0.63, \, 1.38)\\
\bottomrule
\end{tabularx}
\end{table}

\begin{table}[H]
\centering
\caption{Posterior estimates of $\tau(u)$ for correlation $\rho=0.6$ (Assumption 3)  under two scenarios using correlation values $\rho^*_z=0.3,0.6$ (Assumption 4) for $z=0,1$, with 95$\%$ point-wise credible intervals.}\label{estimand2-onedeath}
\begin{tabularx}{\textwidth}{XXX}
\toprule
 Threshold  &  $\rho_z^*=0.3$ &  $\rho_z^*=0.6$ \\
\midrule
 $u=10$ &0.85  \; (0.53, \, 1.24) & 0.76  \; (0.45, \, 1.21) \\
 $u=20$ & 0.92  \; (0.68, \, 1.16) & 0.91   \; (0.63, \, 1.22)\\
 $u=30$  & 0.96  \; (0.76, \, 1.15) & 0.99   \; (0.71, \, 1.34)\\
 $u=40$ & 0.94  \; (0.76, \, 1.12)& 0.93   \; (0.64, \, 1.29) \\
 \bottomrule
\end{tabularx}
\end{table}

\subsection{Results for the Two Terminal Events}

For posterior inference we again used the EDPM truncation approximation (with $N=10$ and $M=8$), based on 50000 MCMC iterations with a burn in period of 25000 iterations. The posterior mean estimates of the causal estimand $\tau(u)$ (defined in \ref{new-estimand}) for the two terminal events are shown in Tables \ref{2Death-table1} through \ref{2Death-table4}. Across correlation values $\rho=0.3,0.6$ (Assumption 3), $\rho^*_z=0.3,0.6$, $z=0,1$ (Assumption 4) and $\rho^{**}$ (Assumption 6), the estimates consistently suggest an increased risk of the non-terminal event under treatment by time $u$ (where $u$ denotes the follow‑up time in years, here evaluated at 10, 20, 30, and 40 years), with all point estimates exceeding 1. The $95\%$ credible intervals exclude 1, indicating strong evidence that individuals receiving treatment have a higher risk of experiencing the non-terminal event by year $u$, among those who survive past year $u$ and ultimately die from CVD. The similarity of results across correlation assumptions implies robustness to these modeling choices.

\begin{table}[H]
\centering
\caption{Posterior mean estimates of $\tau(u)$ under correlation values $\rho=0.3$ (Assumption 3), $\rho^*_{z}=0.3$, \, $z=0,1$ (Assumptions 4) and  $\rho^{**}=0.3,0.6$ ( Assumption 6), with $95\%$ point-wise credible intervals.}\label{2Death-table1}
\begin{tabularx}{\textwidth}{XXX}
\toprule
Threshold 
  & \shortstack{$\rho= 0.3$\\ $\rho_z^*= 0.3$\\$\rho^{**} = 0.3$} 
  & \shortstack{$\rho=  0.3$\\ $\rho_z^*= 0.3$\\$\rho^{**} = 0.6$} \\
\midrule
$u = 10$ & 1.60 \; (1.10, 2.25) & 1.60 \; (1.09, 2.24)\\ 
$u = 20$ & 1.32 \; (1.04, 1.63) & 1.32 \;(1.05, 1.63) \\ 
$u = 30$ & 1.22 \; (1.04, 1.41) & 1.22 \; (1.04, 1.40)\\ 
$u = 40$ & 1.16 \; (1.04, 1.28) & 1.16 \; (1.04, 1.29) \\ 
\bottomrule
\end{tabularx}
\end{table}

\begin{table}[H]
\centering
\caption{Posterior mean estimates of $\tau(u)$ under correlation values $\rho=0.3$ (Assumption 3), $\rho^*_{z}=0.6$, \, $z=0,1$   (Assumption 4) and  $\rho^{**}=0.3, 0.6$ (Assumption 6), with $95\%$ point-wise credible intervals.}\label{2Death-table2}
\begin{tabularx}{\textwidth}{XXX}
\toprule
Threshold 
  & \shortstack{$\rho= 0.3$\\ $\rho_z^*= 0.6$\\$\rho^{**} = 0.3$} 
  & \shortstack{$\rho=  0.3$\\ $\rho_z^*= 0.6$\\$\rho^{**} = 0.6$} \\
\midrule
$u = 10$ & 1.60 \; (1.08, 2.24) & 1.60 \; (1.09, 2.24)\\ 
$u = 20$ & 1.33 \; (1.05, 1.63) & 1.33 \;(1.04, 1.62) \\ 
$u = 30$ & 1.22 \; (1.04, 1.40) & 1.22 \; (1.04, 1.41)\\ 
$u = 40$ & 1.16 \; (1.04, 1.28) & 1.16 \; (1.04, 1.29) \\ 
\bottomrule
\end{tabularx}
\end{table}

\begin{table}[H]
\centering
\caption{Posterior mean estimates of $\tau(u)$ under correlation values $\rho=0.6$ (Assumption 3), $\rho^*_{z}=0.3$, \,$z=0,1$ (Assumption 4) and $\rho^{**}=0.3=0.6$ (Assumption 6), with $95\%$ point-wise credible intervals.}\label{2Death-table3}
\begin{tabularx}{\textwidth}{XXX}
\toprule
Threshold 
  & \shortstack{$\rho= 0.6$\\ $\rho_z^*= 0.3$\\$\rho^{**} = 0.3$} 
  & \shortstack{$\rho=  0.6$\\ $\rho_z^*= 0.3$\\$\rho^{**} = 0.6$} \\
\midrule
$u = 10$ & 1.60 \; (1.09, 2.23)& 1.60 \; (1.09, 2.23) \\ 
$u = 20$ &  1.32 \; (1.04, 1.62) & 1.32 \; (1.04, 1.64)\\ 
$u = 30$ & 1.22 \; (1.04, 1.40)  & 1.22 \; (1.04, 1.40) \\ 
$u = 40$ & 1.16 \; (1.03, 1.29) &
 1.16 \; (1.04, 1.29) \\ 
\bottomrule
\end{tabularx}
\end{table}

\begin{table}[H]
\centering
\caption{Posterior mean estimates of $\tau(u)$ under correlation values $\rho=0.6$ (Assumption 3), $\rho^*_{z}=0.6$, \, $z=0,1$ (Assumption 4)  and $\rho^{**}=0.3,0.6$ (Assumption 6), with $95\%$ point-wise credible intervals.}\label{2Death-table4}
\begin{tabularx}{\textwidth}{XXX}
\toprule
Threshold 
  & \shortstack{$\rho= 0.6$\\ $\rho_z^*= 0.6$\\$\rho^{**} = 0.3$} 
  & \shortstack{$\rho=  0.6$\\ $\rho_z^*= 0.6$\\$\rho^{**} = 0.6$} \\
\midrule
$u = 10$ & 1.60 \; (1.09, 2.25)& 1.60  \; (1.09, 2.24) \\ 
$u = 20$ &  1.32 \; (1.05, 1.63) & 1.32  \; (1.05, 1.62)\\ 
$u = 30$ & 1.22 \; (1.04, 1.40)  & 1.22  \; (1.04, 1.40) \\ 
$u = 40$ &  1.16 \; (1.03, 1.29) & 1.16  \; (1.03, 1.28) \\ 
\bottomrule
\end{tabularx}
\end{table}

\section{Discussion}
 
In this project, we developed a framework for causal inference in semi-competing risks settings, where a non-terminal event (e.g., disease progression) may be censored by one or more terminal events (e.g., death). We proposed a principal stratification-based causal estimand that characterizes the treatment effect on the timing of the non-terminal event, conditional on survival beyond a prespecified time $u$. This formulation was extended from a single terminal event to a setting involving two distinct causes of death, thereby providing a more nuanced perspective on progression when competing terminal events are present.

To identify the estimand, we laid out a set of causal assumptions under a potential outcomes framework and addressed the complexities that arise in the presence of semi-competing risks. For modeling the joint distribution of time-to-events, we employed two flexible approaches: a  vine factorization and a BNP model (EDPM) and copulas for unidentified pairwise conditional distributions. These methods allowed us to capture complex dependence structures and to avoid strong parametric assumptions. Posterior distributions for the causal estimand were obtained via MCMC, and we summarized inference through the posterior mean and credible intervals.

The results demonstrated the utility of both modeling approaches for estimating the causal estimand. Moving forward, a key objective is to compare these Bayesian estimators with standard parametric alternatives, such as the parametric AFT model, to evaluate the estimation accuracy and robutstness of our observed data model. 

There remain important directions for future research. One possible extension is to refine or relax some of the causal identification assumptions, particularly those related to conditional independence in Assumption 4 (Section \ref{Sec. Assumptions two deaths}), for two terminal events setting. Additionally, the method could be extended to estimate heterogeneous treatment effects across patient subgroups by only integrating over a subset of the baseline covariates.

This work contributes to the growing body of literature on causal inference under semi-competing risks. 

\bibliography{referenceFile}
\newpage
\section*{Appendix A: Post-processing steps for estimation of causal estimand (G-computation) for the semi-competing risks model with one terminal event}\label{appendix A}

\begin{enumerate}   
 \item  Let the index $k\in\{1,\dots,N\}$ refer to an upper level cluster, and for  index $j\in\{1,\dots,M\}$ refer to a lower level cluster. Extract posterior draws from the MCMC output for the following quantities:
        \begin{itemize}
        \item Cluster-specific weights, $\gamma_k$ and $\gamma_{jk}$.
          \item Regression coefficients for death and progression, $\beta_{D}$ and $\beta_{P}$, and error variances, $\sigma^2_{D}$ and $\sigma^2_{P}$ for each upper cluster $k$.
          \item  Treatment‐effect coefficient, $\beta_Z$ for each upper cluster $k$.
          \item Covariate distribution parameters corresponding to each upper cluster \(k\) and lower cluster \(j\), 
          including \(\lambda_{jk}\) and \(\tau_{jk}\) for continuous covariates, and \(\psi_{jk}\) for binary covariates.
          \end{itemize}
\item
For each posterior sample, perform $n_{\text{MC}}$ Monte Carlo replicates as follows: Draw an upper cluster \(k\)  and a lower cluster $j \mid k$ using the posterior sample of the weights, 
\begin{align*}
 k \;&\sim\;\mathrm{Categorical}(\gamma_1,\ldots,\gamma_N)\\ 
  j\mid k \;& \sim\;\mathrm{Categorical}(\gamma_{1\mid k},\ldots,\gamma_{M\mid k})  
\end{align*}

\item For the selected mixture component indexed by $(k,j)$, sample the subject’s covariates $X$.
Each continuous covariate is drawn from a Log‑Normal distribution whose parameters depend on $(j,k)$, while each binary covariate is drawn from a Bernoulli distribution:
\begin{align*}
  X_{\mathrm{cont}} 
    &\sim \mathrm{Lognormal}\bigl(\lambda_{jk},\,\sqrt{\tau_{jk}}\bigr),
    &&\text{for each continuous predictor,}\\
  X_{\mathrm{bin}}
    &\sim \mathrm{Bernoulli}\bigl(\psi_{jk}\bigr),
    &&\text{for each binary covariate.}
\end{align*}
We collect all covariates (including the intercept) into a vector
$$X = \begin{pmatrix} 1, & X_{\text{AGE}}, & X_{\text{SBP}}, & X_{\text{CHL}}, & X_{\text{HDL}}, & X_{\text{BMI}}, & X_{\text{SMOKER}}, & X_{\text{GENDER}}, & X_{\text{DIAB}} \end{pmatrix}^\top,$$

\item For each treatment $z=0,1$, compute the normalized weights
$$
w_k(X,z)
=\frac{\gamma_k \sum_{j=1}^M \gamma_{j\mid k}\,P\bigl(X,z\mid\omega_{j\mid k}\bigr)}
{\sum_{h=1}^N \gamma_h \sum_{j=1}^M \gamma_{j\mid h}\,P\bigl(X,z\mid\omega_{j\mid h}\bigr)},
\quad k=1,\dots,N.
$$

\item Simulate \((Y_D^1,Y_D^0)\) from the Gaussian copula $G_\mathbf{x}$, with marginal CDFs $G_\mathbf{x}^1$ and $G_\mathbf{x}^0$,
\begin{equation*}
 G_\mathbf{x}(t,w ;\rho)= \Phi_{2,\rho}\Big[\Phi^{-1}\left\{G_\mathbf{x}^1(t)\right\}, \Phi^{-1}\left\{G_\mathbf{x}^0(w)\right\}\Big]   
\end{equation*}
\noindent In the steps that follow, we begin by sampling on the observed log-survival distribution, but we subsequently transform the outcomes to the standard normal latent scale and remain there throughout the procedure. In particular, we do not transform back to the original outcome scale. 

\begin{enumerate}
\item First, draw a cluster $k'(X,1)$
$$k'(X,1) \sim \mathrm{Categorical}\bigl(w_1(X,1), \dots, w_N(X,1)\bigr).$$ 

\item Extract the regression coefficients $\beta^{(k'(X,1))}_{D}$, $\beta_Z^{(k'(X,1))}$, and error variance $\sigma^{2(k'(X,1))}_{D}$,  and compute  the conditional mean:
\begin{align}
\mu^{(k')}_{D}(X,1) &= X\,\beta_{D}^{(k'(X,1))} + \beta_{Z}^{(k'(X,1))}.
\end{align}
Conditional on this selected component \(k'\), draw the log-survival outcome
$$Y_D^1\sim \mathcal{N}(\mu^{(k')}_{D}(X,1),\; \sigma^{2(k')}_{D}(X,1)),$$ 
 and denote the realized value by $t = Y_D^1$. 

\vspace{6pt}  
\noindent This two‐step procedure  produces the mixture CDF
$$
G_\mathbf{x}^1(t)=\Pr\bigl(Y_D^1\le t\mid X\bigr)
=\sum_{k=1}^N w_k(X,1)\,
\Phi\!\Bigl(\tfrac{t - \mu_{D}^{(k)}(X,1)}{\sigma_{D}^{(k)}}\Bigr)\,. 
$$

where $ w_k(X,1)
=\Pr\bigl(k\mid X,Z=1\bigr)$,
\quad $\mu_D^{(k)}(X,1)=X\,\beta_{D}^{(k(X,1))} + \beta_{Z}^{(k(X,1))}$,
\; and \;
$\sigma_D^{(k)}$ is the standard deviation for death associated with the \,$k$th mixture component. 

\item For the realized value $t$ compute the latent Gaussian score \;    $v_D^1\;=\;\Phi^{-1}\bigl(G_\mathbf{x}^1(t)\bigr)$. \\

Under the Gaussian‐copula assumption sample the conditional latent variable $V_D^0 \mid V_D^1$   
 $$V_D^0 \mid V_D^1=v_D^1 \;\sim\; \mathcal{N}\bigl(\rho \,v_D^1,\; 1-\rho^2\bigr).$$
  \end{enumerate}

\item  Define the unconditional threshold \(u_D^1\),  such that $\Pr(V_D^1 \le u_D^1) \;=\; \Pr(Y_D^1\le \log(u))=G_\mathbf{x}^1(\log(u))$, 
$$u_D^1 \;=\; \Phi^{-1}\bigl(G_\mathbf{x}^1(\log(u))\bigr).$$

Define the threshold \(u_D^0\)
$$u_D^0 \;=\; \rho \, v_D^1 \,+\, \sqrt{1-\rho^2}\;\Phi^{-1}\bigl(G_\mathbf{x}^0(\log(u)\bigr).$$
where
$$ G_\mathbf{x}^0( \log(u))
\;=\;
\sum_{k=1}^{N} w_k(X,0) 
\;\Phi\!\Bigl(\frac{\log(u) -  \mu^{(k)}_{D}(X,0)}{\sigma^{(k)}_{D}(X,0)}\Bigr).$$

\item Test the event $\;Y_D^1 \geq \log(u) \; \text{and} \;  Y_D^0 \geq \log(u)$  by the equivalent copula conditions $\;V_D^1 \geq u_D^1$ and $ V_D^0 \geq u_D^0$.

 \item A simulated subject in step 5, survived if \; $V_D^1 \geq u_D^1 \quad \text{and} \quad V_D^0 \geq u_D^0$.\\
  
  \item Conditional on survival, the two latent progression outcomes (non-terminal HF) are simulated. The conditional joint distribution function of $(Y_P^z, Y_D^{1-z})$ given $(Y_D^z =t, X = x)$,  $H^z_x$ for $z=0,1$ follows a Gaussian
copula model, i.e.,
\begin{equation*}
 H_\mathbf{x}^z(s,r \mid t ;\rho^*_z)= \Phi_{2,\rho^*_z}\Big[\Phi^{-1}\left\{\Pr(Y_p^z \leq s \mid Y_D^z=t]\right\}, \Phi^{-1}\left\{\Pr(Y_D^{1-z} \leq r \mid Y_D^z = t]\right\}\Big].  
\end{equation*}

\noindent The joint distribution $H_\mathbf{x}^z$ (for $z=0,1$) involves the conditional distribution of \,$Y_D^{1-z} \mid Y_D^z$, which is modeled via the Gaussian copula (Assumption 4 in Section \ref{Assumptions One death}). Since the dependence is closed form in the latent space, we express this conditional distribution through the corresponding latent variables. From earlier steps, we know the conditional distribution of $V_D^{1-z} \mid V_D^z$, the first copula edge,
$$V_D^{1-z} \mid V_D^z=v_D^z \;\sim\; \mathcal{N}\bigl(\rho \, v_D^z,\; 1-\rho^2\bigr).$$
Given the realized pair $(v_D^z, v_D^{1-z})$, we compute the conditional standard normal score:
$$ v_{\text{std}}^{1-z}=\Phi^{-1}\bigl(\Pr(V_D^{1-z} \leq v_D^{1-z} \mid V_D^z = v_D^z)\bigr)
\;=\; \Phi^{-1}\!\Bigl(\Phi\!\Bigl(\frac{v_D^{1-z} - \rho\,v_D^z}{\sqrt{1-\rho^2}}\Bigr)\Bigr)=\frac{v_D^{1-z} - \rho\,v_D^z}{\sqrt{1-\rho^2}}.
$$
This transformed score $v_{\text{std}}^{1-z}$ is then used as the conditioning value to sample the latent variable of the second copula edge  $V_P^z \mid V_D^z$,
$$V_P^z \mid V_D^z  \sim \mathcal{N}\left(\rho_z^*\; v_{\text{std}}^{1-z},\; \sqrt{1-(\rho_z^*)^2}\right).$$

\item  Define thresholds \(u_{P}^z\),  
$$ u_{P}^z \;=\; \rho^*_z \; v_{\text{std}}^{1-z}\; + \; \sqrt{1- (\rho^*_z)^2}\;\Phi^{-1}\bigl(\Pr(Y_P^z\le \log(u))\bigr), \quad z=0,1,$$
with the CDF calculated as 
 $$ \Pr(Y_P^z \leq \log(u)) \;=\;
\sum_{k=1}^{N} w_k(X,z) 
\;\Phi\!\Bigl(\frac{\log(u) -  \mu^{(k)}_{P}(X,z)}{\sigma^{(k)}_{P}(X,z)}\Bigr), 
$$
where $ w_k(X,z)
=\Pr\bigl(k\mid X,Z=z\bigr)$,
\quad $\mu_P^{(k)}(X,z)=X\,\beta_{P}^{(k(X,z))} + z\,\beta_{Z}^{(k(X,z))}$,
\; and \;
$\sigma_P^{(k)}$ is the standard deviation of progression associated with the \,$k$th mixture component. 

\item Test the event $ \;Y_P^z < \log(u)$ by the equivalent copula condition $\;V_P^z < u_P^z$.\\

\item Monte Carlo integration and estimand calculation:  
    \begin{itemize}
      \item For each valid simulation (i.e., when the survival criteria are met), event indicators are recorded:
     $$ I_1 = \mathbf{1}\{V_P^1 < u_P^1\} \quad \text{and} \quad I_0 = \mathbf{1}\{V_P^0 < u_P^0\}.
       $$
      \item The numerator and the denominator are defined as the summed counts of events, respectively:
$$\text{num} = \sum I_1, \quad \text{den} = \sum I_0.$$
      \item After obtaining $n_{MC}$ valid draws, the final estimand is computed as: 
    \begin{align*}
    \hat{\tau}(u) &= \frac{\text{num}}{\text{den}}= \frac{ \sum_{m=1}^{n_{\text{MC}}} \mathbf{1}\{{V_P^1 < u_P^1} \mid  V_D^1 \geq u_D^1 , \; V_D^0 \geq u_D^0 \} }{ \sum_{m=1}^{n_{\text{MC}}} \mathbf{1}\{ V_P^0 < u_P^0 \mid   V_D^1 \geq u_D^1 , \; V_D^0 \geq u_D^0  \}}.
\end{align*}
    \end{itemize}

 \item After computing $\hat{\tau}(u)$ for all $n_{\text{posterior}}=1000$ samples, we have 1000 posterior samples of the estimands. We obtain the final estimate as:
\begin{align*}
    \mathbb{E}[\tau(u)] \approx \frac{1}{n_{\text{posterior}}} \sum_{i=1}^{n_{\text{posterior}}} \hat{\tau}_i(u).
\end{align*}
\end{enumerate}
\newpage
\section*{Appendix B: Specifications for the Semi-Competing risks model with One Terminal Event}\label{appendix model specification}

\noindent We model the logarithm of the death time $Y_{D}$ and the HF time $Y_{P}$ with each mixture component of the EDPM using AFT models:
\begin{align*}
\log(Y_{D}) &= X\beta_{D} + \varepsilon_{D}, 
& \varepsilon_{D} &\overset{\text{iid}}{\sim} N(0,\sigma_{D}^{2}),\\
\log(Y_{P}) &= X\beta_{P} + \varepsilon_{P}, 
& \varepsilon_{P} &\overset{\text{iid}}{\sim} N(0,\sigma_{P}^{2}).
\end{align*}

\noindent where $X\in\mathbb{R}^{n\times p}$ is the design matrix (including intercept), containing both continuous and binary covariates.

\subsection{Priors}

\subsubsection{Regression and Scale Parameters (Death and HF)}  
Let $\hat\beta_D,\hat\sigma_D^2$ and $\hat\beta_{P},\hat\sigma_{P}^2$ be the maximum likelihood estimates and the error variance (scale) estimates. We assign
\begin{align*}
\beta_{D} &\sim N\!\Bigl(\hat\beta_{D},\;\frac{n_{\mathrm{death}}}{5}\,\mathrm{diag}(\hat\Sigma_{\beta_{D}})\Bigr), 
& 
\sigma_{D}^{2} &\sim \mathrm{Inv\text{-}Gamma}\bigl(3,\;\hat\sigma_{D}^{2}\bigr),\\[3pt]
\beta_{P} &\sim N\!\Bigl(\hat\beta_{P},\;\frac{n_{P}}{5}\,\mathrm{diag}(\hat\Sigma_{\beta_{P}})\Bigr), 
& 
\sigma_{P}^{2} &\sim \mathrm{Inv\text{-}Gamma}\bigl(3,\;\hat\sigma_{P}^{2}\bigr).
\end{align*}
Here $\hat\Sigma_{\beta}$ denotes the estimated covariance of $\hat\beta$, and $n_D,n_{P}$ are the numbers of uncensored events in each model.

\subsubsection{Covariates}  
For each continuous covariate $X_h$, let $\bar X_h$ and $\hat\sigma_{X_h}^2$ be its sample mean and variance.  We set
$$ X_h \;\sim\; N(\lambda_h,\tau_h),
\quad
\lambda_h\;\sim\;N\Bigl(\bar X_h,\;\frac{n}{5}\,\hat\sigma_{X_h}^2\Bigr),
\quad
\tau_h\;\sim\;\mathrm{Inv\text{-}Gamma}\bigl(3,\;2\,\hat\sigma_{X_h}^2\bigr).$$

For each binary covariate $X_k$, let $\bar X_k$ be its observed proportion.  We set
$$
X_k\;\sim\;\mathrm{Bernoulli}(\psi_k),
\quad
\psi_k\;\sim\;\mathrm{Beta}\bigl(10\,\bar X_k,\;10\,(1-\bar X_k)\bigr).
$$
\newpage 
\section*{Appendix C: Post-processing steps for estimation of causal estimand for the semi-competing risks model with two terminal events}\label{appendix C}

\begin{enumerate}   
 \item  Let the index $k\in\{1,\dots,N\}$ refer to an upper level cluster, and for  index $j\in\{1,\dots,M\}$ refer to a lower level cluster. Extract posterior draws from the MCMC output for the following quantities:
        \begin{itemize}
        \item Cluster-specific weights, $\gamma_k$ and $\gamma_{jk}$.
          \item Regression coefficients for death and progression, $\beta_{D_1}$, $\beta_{D_2}$ and $\beta_{P}$, and error variances, $\sigma^2_{D_1}$ , $\sigma^2_{D_2}$ and $\sigma^2_{P}$ for each upper cluster $k$.
          \item  Treatment‐effect coefficient, $\beta_Z$ for each upper cluster $k$.
          \item Covariate distribution parameters corresponding to each upper cluster \(k\) and lower cluster \(j\), 
          including \(\lambda_{jk}\) and \(\tau_{jk}\) for continuous covariates, and \(\psi_{jk}\) for binary covariates.
        \end{itemize}
\item
For each posterior sample, perform $n_{\text{MC}}$ Monte Carlo replicates as follows: Draw an upper cluster \(k\)  and a lower cluster $j \mid k$ using the posterior sample of the weights, 
\begin{align*}
 k \;&\sim\;\mathrm{Categorical}(\gamma_1,\ldots,\gamma_N)\\ 
  j\mid k \;& \sim\;\mathrm{Categorical}(\gamma_{1\mid k},\ldots,\gamma_{M\mid k})  
\end{align*}

\item For the selected mixture component indexed by $(k,j)$, sample the subject’s covariates $X$.
Each continuous covariate is drawn from a Log‑Normal distribution whose parameters depend on $(j,k)$, while each binary covariate is drawn from a Bernoulli distribution:
\begin{align*}
  X_{\mathrm{cont}} 
    &\sim \mathrm{Lognormal}\bigl(\lambda_{jk},\,\sqrt{\tau_{jk}}\bigr),
    &&\text{for each continuous predictor,}\\
  X_{\mathrm{bin}}
    &\sim \mathrm{Bernoulli}\bigl(\psi_{jk}\bigr),
    &&\text{for each binary covariate.}
\end{align*}
We collect all covariates (including the intercept) into a vector
$$X = \begin{pmatrix} 1, & X_{\text{AGE}}, & X_{\text{SBP}}, & X_{\text{CHL}}, & X_{\text{HDL}}, & X_{\text{BMI}}, & X_{\text{SMOKER}}, & X_{\text{GENDER}}, & X_{\text{DIAB}} \end{pmatrix}^\top,$$

\item For $z=0,1$, compute the normalized weights
$$w_k(X,z) = \dfrac{
\gamma_k \sum_{j=1}^M \gamma_{j\mid k} P(X,z \mid \omega_{j\mid k})}{\sum _{h=1}^N \gamma_h \sum_{j=1}^M \gamma_{j\mid h} P(X,z \mid \omega_{j\mid h})},  \quad k=1, \dots, N.$$

\item Simulate \((Y_{D_2}^1,Y_{D_2}^0)\) from the Gaussian copula $H_\mathbf{x}$,  
 \begin{equation*}
 H_\mathbf{x}(t,w)= \Phi_{2,\rho}\Big[\Phi^{-1}\left\{ \Pr(Y_{D_2}^1 \leq t)\right\}, \Phi^{-1}\left\{ \Pr(Y_{D_2}^{0} \leq w)\right\}\Big].  
\end{equation*} 
\noindent In the steps that follow, we begin by sampling on the observed log-survival distribution, but we subsequently transform the outcomes to the standard normal latent scale and remain there throughout the procedure. In particular, we do not transform back to the original outcome scale.

\begin{enumerate}
\item First, draw the cluster $k'(X,z)$
$$k'(X,1) \sim \mathrm{Categorical}\bigl(w_1(X,1), \dots, w_N(X,1)\bigr).$$ 
\item Extract the regression coefficients $\beta^{(k'(X,1))}_{D_2}$, \; $\beta_Z^{(k'(X,1))}$, and error variance $\sigma^{2(k'(X,1))}_{D_2}$,\; and compute  the conditional mean:
\begin{align}
\mu^{(k')}_{D_2}(X,1) &= X\,\beta_{D_2}^{(k'(X,1))} + \beta_{Z}^{(k'(X,1))}.
\end{align}

Conditional on this selected component $k'$, draw the log-survival outcome 
$$Y_{D_2}^1\sim \mathcal{N}(\mu^{(k')}_{D_2}(X,1),\; \sigma^{2(k')}_{D_2}(X,1)),$$ and denote the realized value by $t = Y_{D_2}^1$. \\
 
\noindent This two‐step procedure  produces the mixture CDF
$$
\Pr\bigl(Y_{D_2}^1\le t\mid X\bigr)
=\sum_{k=1}^N w_k(X,1)\,
\Phi\!\Bigl(\tfrac{t - \mu_{D_2}^{(k)}(X,1)}{\sigma_{D_2}^{(k)}}\Bigr), 
$$
where $ w_k(X,1)
=\Pr\bigl(k\mid X,Z=1\bigr)$,
\quad $\mu_{D_2}^{(k)}(X,1)=X\,\beta_{D_2}^{(k(X,1))} + \beta_{Z}^{(k(X,1))}$,
\; and \;
$\sigma_{D_2}^{(k)}$ is the standard deviation for death due to non-CVD associated with the \,$k$th mixture component. 

\item For the realized value $t$ compute the latent Gaussian score \;    $v_{D_2}^1\;=\;\Phi^{-1}\bigl(\Pr(Y_{D_2}^1 \leq t)\bigr)$.\\
Under the Gaussian‐copula assumption sample the corresponding conditional latent variable  
 $$V_{D_2}^0 \mid V_{D_2}^1=v_{D_2}^1 \;\sim\; \mathcal{N}\bigl(\rho \,v_{D_2}^1,\; 1-\rho^2\bigr).$$
  \end{enumerate}

  \item The conditional joint distribution function of $(Y_{D_1}^z, Y_{D_2}^{1-z})$ given $(Y_{D_2}^z =t, X = x)$,  $J^z_x$ for $z=0,1$ follows a Gaussian
copula model, i.e.,
\begin{equation*}
 J_\mathbf{x}^z(s,w \mid t ;\rho^*_z)= \Phi_{2,\rho^*_z}\Big[\Phi^{-1}\left\{\Pr(Y_{D_1}^z \leq s \mid Y_{D_2}^z=t]\right\}, \Phi^{-1}\left\{\Pr(Y_{D_2}^{1-z} \leq w \mid Y_{D_2}^z = t]\right\}\Big].  
\end{equation*}

\noindent The quantity $J_\mathbf{x}^z$ (for $z=0,1$) involves the conditional distribution of \,$Y_{D_2}^{1-z} \mid Y_{D_2}^z$, which was modeled via the Gaussian copula $H_\mathbf{x}$. Since the dependence is closed form in the latent space, we express this conditional distribution through the corresponding latent variables. From earlier steps sampling using $H_\mathbf{x}$, we know the conditional distribution of $V_{D_2}^{1-z} \mid V_{D_2}^z$, the first copula edge,
$$V_{D_2}^{1-z} \mid V_{D_2}^z=v_{D_2}^z \;\sim\; \mathcal{N}\bigl(\rho \, v_{D_2}^z,\; 1-\rho^2\bigr).$$
Given the realized pair $(v_{D_2}^z, v_{D_2}^{1-z})$, we compute the conditional standard normal scores:
$$ v_{\text{std}}^{1-z}=\Phi^{-1}\bigl(\Pr(V_{D_2}^{1-z} \leq v_{D_2}^{1-z} \mid V_{D_2}^z = v_{D_2}^z)\bigr)
\;=\; \Phi^{-1}\!\Bigl(\Phi\!\Bigl(\frac{v_{D_2}^{1-z} - \rho\,v_{D_2}^z}{\sqrt{1-\rho^2}}\Bigr)\Bigr)=\frac{v_{D_2}^{1-z} - \rho\,v_{D_2}^z}{\sqrt{1-\rho^2}}.
$$
Therefore, $V_{D_1}^z \mid V_{D_2}^z$  using the transformed score $v_{\text{std}}^{1-z}$ in the Gaussian copula $J_\mathbf{x}^z$ sequence:
$$V_{D_1}^z \mid V_{D_2}^z  \sim \mathcal{N}\left(\rho^*_z\; v_{\text{std}}^{1-z},\; \sqrt{1-\rho_2^2}\right).$$

For $z=1$ we sample $V_{D_1}^1 \mid V_{D_2}^1 \sim \mathcal{N}\left(\rho_z^*\; v_{\text{std}}^0,\; \sqrt{1-(\rho_z^*)^2}\right)$  and denote the realized value by $v_{D_1}^1$.\\


\item To sample $V_{D_1}^0 \mid V_{D_2}^0,\,V_{D_2}^1$ we use the copula $L_\mathbf{x}$. The conditional joint distribution function of $(Y_{D_1}^1, Y_{D_1}^0)$ given $(Y_{D_2}^1=t, Y_{D_2}^0=w, \mathbf{X}=\mathbf{x})$, $L_\mathbf{x}$, follows the copula model
\small{
\begin{equation*}
 L_\mathbf{x}(s,v \mid t, w)= \Phi_{2,\rho^{**}}\Big[\Phi^{-1}\left\{ \Pr(Y_{D_1}^1 \leq s \mid Y_{D_2}^1=t, Y_{D_2}^0=w)\right\}, \Phi^{-1}\left\{ \Pr(Y_{D_1}^0 \leq v  \mid Y_{D_2}^1=t, Y_{D_2}^0=w)\right\}\Big].  
\end{equation*}}
Specifically, we draw
$$ V_{D_1}^0 \mid V_{D_1}^0, V_{D_1}^0
\;\sim\;
\mathcal{N}\Bigl(
\rho^{**}\;v_{\text{std}}^{1}\,,\; 1-(\rho^{**})^{2}
\Bigr).$$

\item Define the thresholds $u_{D_1}^z$, $z=0,1$ 
\begin{align*}
u_{D_1}^1 \;&=\; \rho^*_z \; v_{\text{std}}^{0}\; + \; \sqrt{1- (\rho^*_z)^2}\;\Phi^{-1}\bigl(\Pr(Y_{D_1}^1\le \log(u))\bigr),\\
u_{D_1}^0 \;&=\; \rho^{**} \; v_{\text{std}}^{1}\; + \; \sqrt{1- (\rho^{**})^2}\;\Phi^{-1}\bigl(\Pr(Y_{D_1}^0\le \log(u))\bigr),
\end{align*}

with the CDF calculated as 
 $$ \Pr(Y_{D_1}^z \leq \log(u)) \;=\;
\sum_{k=1}^{N} w_k(X,z) 
\;\Phi\!\Bigl(\frac{\log(u) -  \mu^{(k)}_{D_1}(X,z)}{\sigma^{(k)}_{D_1}(X,z)}\Bigr), \quad z=0,1, 
$$
where $ w_k(X,z)
=\Pr\bigl(k\mid X,Z=z\bigr)$,
\quad $\mu_{D_1}^{(k)}(X,z)=X\,\beta_{D_1}^{(k(X,z))} + z\,\beta_{Z}^{(k(X,z))}$,
\; and \;
$\sigma_{D_1}^{(k)}$ is the standard deviation of death due to CVD associated with the \,$k$th mixture component. 

Test the event \( \;Y_{D_1}^z \geq \log(u)\) by the equivalent copula condition \(\;v_{D_1}^z \geq u_{D_1}^z\), for $z=0,1$.\\

\item Consider simulated subject who survive beyond time $u$ and under either medication status $z\in\{0,1\}$, experience death due to CVD ``before" death due to other causes. Formally, \;  $V_{D_2}^1 \geq V_{D_1}^1 \geq u_{D_1}^1$ and $\quad V_{D_2}^0 \geq V_{D_1}^0 \geq u_{D_1}^0$.

\item For each treatment $z=0,1$, sample the latent log-times for HF

\begin{enumerate}
    \item First, conditional on the selected component $k'(X,z)$,  extract the  regression coefficients $\beta_{P}^{(k'(X,z))}$ and error variances $\sigma^2_P(X,z))$. Compute the mean
    $$\mu^{(k')}_{P}(X,z) = X\,\beta_{P}^{(k'(X,z))} + z\,\beta_{Z}^{(k'(X,z))}.$$
    
    \item Sample $$ Y_P^z \, \sim \mathcal{N} (\mu^{(k')}_P(X,z), \sigma^{2(k')}_P(X,z)),$$ 
    and denote the realized value by $r^z=Y_P^z$.

\noindent This two‐step procedure  produces the mixture CDF
$$
\Pr\bigl(Y_{P}^z\le t\mid X\bigr)
=\sum_{k=1}^N w_k(X,z)\,
\Phi\!\Bigl(\tfrac{t - \mu_{P}^{(k)}(X,z)}{\sigma_{P}^{(k)}}\Bigr), 
$$
where $ w_k(X,1)
=\Pr\bigl(k\mid X,Z=1\bigr)$,
\quad $\mu_{P}^{(k)}(X,1)=X\,\beta_{P}^{(k(X,1))} + z \, \beta_{Z}^{(k(X,1))}$,
\; and \;
$\sigma_{P}^{(k)}$ is the standard deviation for progression time associated to the \,$k$th mixture component. 
\item Let $V_P^z$ be the latent Gaussian variable for $Y_P^z$ and for the realized value $r^z$ compute the corresponding latent Gaussian scores $v_P^z =\Phi^{-1}(\Pr(Y_P^z \leq r^z))$, $z=0,1$.
\end{enumerate} 

\item For $z=0,1$ define thresholds $u_P^z$
$$u_P^z= \Phi^{-1}(\Pr(Y_P^z \leq \log(u) )).$$

Test the event $\;Y_P^z < \log(u)$ by the equivalent copula condition $\;V_P^z < u_P^z$.\\

\item Monte Carlo integration and estimand calculation:  
    \begin{itemize}
      \item For each valid simulation (i.e, when the simulated subject survive beyond time $u$ and under either medication status $z\in\{0,1\}$, experience death due to CVD ``before" death due to other causes), event indicators are recorded:
     $$ I_1 = \mathbf{1}\{V_P^1 < u_P^1\} \quad \text{and} \quad I_0 = \mathbf{1}\{V_P^0 < u_P^0\}.
       $$
      \item The numerator and the denominator are defined as the summed counts of events, respectively:
$$\text{num} = \sum I_1, \quad \text{den} = \sum I_0.$$
      \item After obtaining $n_{MC}$ valid draws, the final estimand is computed as: 

    \begin{align*}
    \hat{\tau}(u) &= \frac{\text{num}}{\text{den}}= \frac{ \sum_{m=1}^{n_{\text{MC}}} \mathbf{1}\{{V_P^1 < u_P^1} \mid  V_{D_2}^1 \geq V_{D_1}^1 \geq u_{D_1}^1 , \; V_{D_2}^0 \geq V_{D_1}^0 \geq u_{D_1}^0 \} }{ \sum_{m=1}^{n_{\text{MC}}} \mathbf{1}\{ V_P^0 < u_P^0 \mid  V_{D_2}^1 \geq V_{D_1}^1 \geq u_{D_1}^1 , \; V_{D_2}^0 \geq V_{D_1}^0 \geq u_{D_1}^0  \}}.
\end{align*}
    \end{itemize}

 \item After computing $\hat{\tau}(u)$ for all $n_{\text{posterior}}=1000$ samples, we have 1000 posterior samples of the estimands. We obtain the final estimate as:
\begin{align*}
    \mathbb{E}[\tau(u)] \approx \frac{1}{n_{\text{posterior}}} \sum_{i=1}^{n_{\text{posterior}}} \hat{\tau}_i(u).
\end{align*}
\end{enumerate}

\end{document}